\documentstyle [fleqn,12pt] {article}
\def\mscript#1{\mbox{\scriptsize$#1$}}

\begin{document}
\pagestyle{empty}
\begin{flushright}
hep-th/9708161

\parbox{8cm}{\begin{center} The 3rd RESCEU International Symposium \\
PARTICLE COSMOLOGY \\
The University of Tokyo, November 1997 
\end{center}}
\end{flushright}
\vspace{1cm}
\begin{center}
{\bf Thermofield Dynamics of the Heterotic 
String \\
 --- Thermal Cosmological Constant --- } \\
\vspace{1cm} 
H. Fujisaki,$^*$
S. Sano$^*$\footnote[2] {e-mail: sano@rikkyo.ac.jp} and  
K.Nakagawa$^{**}$\footnote[4] {e-mail: nakagawa@hoshi.ac.jp} 

\vspace{5mm}

$^*$ Department of Physics, Rikkyo University, Tokyo 171\\
$^{**}$Faculty of Pharmaceutical Sciences, Hoshi University, Tokyo 142\\
\vspace{3cm}
{\bf ABSTRACT}
\end{center}

\indent The thermofield dynamics of the $D = 10$ 
heterotic thermal string theory is exemplified at any finite 
temperature through the infrared behaviour of the one-loop 
cosmological constant in proper reference to 
the thermal duality symmetry in association with the 
global phase structure of the thermal string ensemble.    
\newpage
\pagestyle{plain}
\setcounter{page}{1}
\indent Building up thermal string theories based upon the thermofield 
dynamics (TFD) has gradually been endeavoured in leaps 
and bounds.   
In the present 
communication, the TFD algorithm of the $D = 
10$ heterotic thermal string theory is recapitulatively exemplified 
$\grave{a}\: la$ recent publication of ourselves \cite{fujisaki1}, 
\cite{fujisaki2}  
through the infrared 
behaviour of the one-loop cosmological constant in proper respect of 
the thermal duality symmetry as well as 
the thermal stability of modular invariance.  The 
global phase structure of the heterotic thermal string ensemble is 
also touched upon.  

Let us start with the one-loop cosmological constant 
$\Lambda(\beta)$ as follows:
\vspace{5mm}
\begin{equation}
\Lambda (\beta) = \frac{\alpha^\prime}{2} \lim_{\mu^2 \rightarrow 0} {\rm 
Tr} 
\left[ \int_{\infty}^{\mu^2} dm^2 \left( \Delta^\beta_B (p, P; m^2) + 
\Delta^{\beta}_{F} (p, P; m^2) \right) \right]
\end{equation}

\vspace{5mm}
\noindent at any finite temperature $\beta^{-1} = kT$ in the $D = 10$ heterotic
 thermal string theory based upon the TFD algorithm, 
where $\alpha^\prime$ means the slope parameter, $p^\mu$ 
reads loop momentum, $P^I$ lie on the root lattice $\Gamma_8 \times \Gamma_8$ 
for the exceptional group $E_8 \times E_8$ and the thermal propagator 
$\Delta^\beta_{B[F]} (p, P; m^2)$ of the free closed bosonic [fermionic] 
string is written $\grave{a}\: la$ Leblanc in the form
 
\begin{eqnarray}
\lefteqn{\Delta^\beta_{B[F]}(p, P; m^2) = \int_{-\pi}^{\pi} 
\frac{d\phi}{4\pi} \; {\rm e}^{i\phi \left( N - \alpha - \bar{N} + 
\bar{\alpha} -1/2 \cdot \sum_{I=1}^{16} (P^I)^2 \right) }}
\nonumber \\
& & \times \Biggl( \left[ \raisebox{-1ex}{$\stackrel {\textstyle 
+}{\mscript{[}-\mscript{]}}$} \int_{0}^{1} dx + \frac{1}{2}
\sum_{n=0}^{\infty} \frac{\delta [\alpha^\prime 
/2 \cdot p^2 + \alpha^\prime /2 \cdot m^2 + 2(n - \alpha)]}{{\rm e}^{\beta 
|p_0|} \raisebox{-1ex}{$\stackrel{\textstyle -}{\mscript{[}+\mscript{]}}$}
\; 1} 
\oint_{c} dx \right] \nonumber \\
& & \times x^{\alpha^\prime /2 \cdot p^2 + N - \alpha + \bar{N} - 
\bar{\alpha} + 1/2 \cdot \sum_{I=1}^{16} (P^I)^2 + \alpha^\prime /2 \cdot 
m^2 - 1} \Biggr) \quad ,
\end{eqnarray}

\vspace{5mm}
\noindent where $N$ [$\bar{N}$] denotes the number operator of the right- 
[left-] 
mover, the intercept parameter $\alpha$ [$\bar{\alpha}$] is fixed at
$\alpha = 0$  
$[\bar{\alpha} = 1]$ and the contour $c$ is taken as the unit circle 
around the origin.  The modular parameter integral representation of 
$\Lambda (\beta)$ is then written $\grave{a}\: la$ O'Brien and Tan  
as follows:
\vspace{5mm}
\begin{eqnarray}
\lefteqn{\Lambda (\beta)  =  -8(2\pi \alpha^\prime)^{-D/2} \int_{E} 
\frac{d^2\tau}{2\pi \tau_2^2} \; (2\pi \tau_2)^{-(D-2)/2}\; {\rm 
e}^{2\pi i \bar{\tau}} \left[ 1 + 480 \sum_{m=1}^{\infty} 
\sigma_7(m)\bar{z}^m \right]} \nonumber \\
& & \times \prod_{n=1}^{\infty} (1 - \bar{z}^n)^{-D-14} \left( 
\frac{1 + z^n}{1 - z^n} \right)^{D-2} \sum_{\ell \in Z; {\rm odd}} \exp \left[ 
- \frac{\beta^2}{4\pi \alpha^\prime \tau_2}\; \ell^2 \right] 
\rule{0cm}{1cm} \; ;\quad D = 10 \quad , 
\end{eqnarray}

\vspace{5mm}
\noindent where $\stackrel{[\normalsize{-}]}{\tau} = \tau_1 
\raisebox{-1ex}{$\stackrel{\normalsize{+}}{\mscript{[}\!-\!\mscript{]}}$} 
i\tau_2$, 
$z = x {\rm e}^{i\phi} = {\rm 
e}^{2\pi i \tau}$, $\bar{z} = x{\rm e}^{-i \phi} = {\rm e}^{-2\pi i 
\bar{\tau}}$, $E$ means the half-strip region in the $\tau$ plane, 
{\it i.e.} $-1/2 \leq \tau_1 \leq 1/2$; 
$\tau_2 > 0$.   
The ``$E$-type'' thermal amplitude $\Lambda (\beta)$ 
obtained above is not modular invariant and annoyed with ultraviolet 
divergences for $\beta \leq 
\beta_H = (2 + \sqrt{2}) \pi \sqrt{\alpha^\prime}$, where $\beta_H$ 
reads the inverse Hagedorn temperature of the heterotic thermal 
string.      

Our prime concern is reduced to regularizing the thermal amplitude 
$\Lambda (\beta)$ {\it $\grave{a}$ la} O'Brien and Tan through transforming 
the physical information in 
the ultraviolet region of the half-strip $E$ into the ``new-fashioned'' 
modular invariant amplitude.  Let us postulate the one-loop dual symmetric 
thermal cosmological 
constant $\bar{\Lambda} (\beta; D)$ at any space-time dimension $D$  as an 
integral over the 
fundamental domain $F$ of the modular 
group 
$SL(2, Z)$ as follows:

\begin{eqnarray}
\bar{\Lambda} (\beta; D) & = & \frac{2}{\beta} (2\pi \alpha^\prime)^{-D/2} 
\sum_{(\sigma, \rho)} 
\int_{F} 
\frac{d^2\tau}{2\pi \tau_2^2}\; (2 \pi \tau_2)^{-(D-2)/2}\;  
\bar{z}^{-(D+14)/24} z^{-(D-2)/24}  \nonumber \\
& & \times \left[ 1 + 480 \sum_{m=1}^{\infty} 
\sigma_7 (m) \bar{z}^m \right] \prod_{n=1}^{\infty} (1 - \bar{z}^n)^{-D-14} 
(1 - z^n)^{-D+2} \nonumber \\[5mm] 
& & \times A_{\sigma \rho} (\tau; D) \left[C_\sigma^{(+)}(\bar{\tau}, \tau;
\beta) 
+ \rho C_\sigma^{(-)}(\bar{\tau}, \tau; \beta) \right] \quad ,
\end{eqnarray}

where
\begin{eqnarray}
\left( \begin{array}{c}
A_{+-}(\tau; D) \rule[-2mm]{0mm}{8mm} \\ A_{-+}(\tau; D)
\rule[-2mm]{0mm}{8mm} \\ A_{--}(\tau; D) 
\end{array} \right) 
= 8 \left( \frac{\pi}{4} \right) ^{(D-2)/6} 
\left( \begin{array}{l}
-[\theta_2(0, \tau)/\theta_{1}^{\prime}(0, \tau)^{1/3}]^{(D-2)/2}
\rule[-2mm]{0mm}{8mm} \\
-[\theta_4(0, \tau)/\theta_{1}^{\prime}(0, \tau)^{1/3}]^{(D-2)/2}
\rule[-2mm]{0mm}{8mm} \\
\; [\theta_3(0, \tau)/\theta_{1}^{\prime}(0, \tau)^{1/3}]^{(D-2)/2}
\rule[-2mm]{0mm}{8mm}
\end{array} \right) \quad ,
\end{eqnarray}

\vspace{5mm}
\begin{equation}
C_\sigma^{(\gamma)}(\bar{\tau}, \tau; \beta) = (4\pi^2\alpha^\prime 
\tau_2)^{1/2} \sum_{(p, q)} \exp \left[ - \frac{\pi}{2} \left( 
\frac{\beta^2}{2\pi^2\alpha^\prime} p^2 + \frac{2\pi^2\alpha^\prime}
{\beta^2} q^2 \right) \tau_2 + i\pi pq \tau_1 \right] ,
\end{equation}

\vspace{5mm}
\noindent the signatures $\sigma, \rho$ and $\gamma$ read $\sigma, \rho = +, 
- ; \; -, + ; \; -, -$ and $\gamma = +, -$, respectively, and the summation 
over $p \; [q]$ is restricted by $(-1)^p = \sigma \; [(-1)^q = 
\gamma]$.   
It is almost needless to mention that $\bar{\Lambda}(\beta; D = 10)$ 
is literally identical with $\Lambda (\beta)$.  The thermal amplitude 
$\bar{\Lambda}(\beta; D)$ is manifestly modular invariant and 
free of ultraviolet divergences for any value of $\beta$ and $D$.    
If and only if $D =  10$, in addition, the thermal duality relation $\beta 
\bar{\Lambda}(\beta; D) = 
\tilde{\beta}\bar{\Lambda}(\tilde{\beta}; D)$ is manifestly satisfied 
for the thermal amplitude $\bar{\Lambda}(\beta; D)$, irrespective of 
the value of $\beta$, where $\tilde{\beta} = 2\pi^2 \alpha^\prime 
/\beta$.     

The infrared behaviour of the 
thermal cosmological constant $\bar{\Lambda}(\beta; D)$ 
is asymptotically described as
\begin{eqnarray}
\bar{\Lambda}(\beta; D) & = & \frac{2}{\beta} \; (4\pi^2 \alpha^\prime)^{-D/2} 
\int_{F} d^2\tau \: \tau_2^{-(D + 2)/2} \exp \left [ \pi \tau_2 \cdot \frac{D + 
6}{6} \right ]
\nonumber \\
& & \times \left[ A_{-+}(\tau; D) - A_{--}(\tau; D) \right]
C_-^{(-)}(\bar{\tau},
 \tau; \beta)  \rule{0cm}{1cm} \quad .
\end{eqnarray}

\vspace{5mm}
\noindent The $D = 10$ TFD amplitude 
$\bar{\Lambda}(\beta; D = 10)$ is then infrared divergent for $(2 - \sqrt{2})
\pi \sqrt{\alpha^{\prime}} = \tilde{\beta}_H \leq \beta \leq \beta_H$, where 
$\tilde{\beta}_H$ reads the inverse dual Hagedorn temperature of the 
heterotic thermal string.   
We can therefore define the dimensionally regularized, 
$D = 10$ one-loop dual symmetric thermal cosmological constant 
$\hat{\Lambda}(\beta)$ in the sense of analytic continuation from $D < 2/5$ 
to $D = 10$ by 
\begin{eqnarray}
\lefteqn{\hat{\Lambda}(\beta)  =   -\frac{2}{\beta} (8\pi 
\alpha^\prime)^{-(D-1)/2} \sum_{(p, q)} 
\int_{- \frac{1}{2}}^{\frac{1}{2}} d\tau_1 \exp[i\pi pq \tau_1]} \nonumber 
\\
& & \times \left( \frac{\beta^2}{2\pi^2\alpha^\prime} \: p^2 + 
\frac{2\pi^2\alpha^\prime}
{\beta^2} \: q^2 - 6 - i\varepsilon \right) ^{(D-1)/2} \nonumber \\[5mm]
& & \times \Gamma \left[ - \frac{D - 1}{2}\: ,\; \frac{\pi}{2} \sqrt{1 - 
\tau_1^2} \left( \frac{\beta^2}{2\pi^2 \alpha^\prime} \: p^2 + 
\frac{2\pi^2 
\alpha^\prime}{\beta^2} \: q^2 - 6 - i\varepsilon \right) \right]\; ;\quad 
D = 10 \quad , 
\end{eqnarray}

\vspace{5mm}
\noindent irrespective of the value of $\beta$, where $p, q = \pm 1; \pm 3; 
\pm 5; \cdots$, and $\Gamma$ is the 
incomplete gamma function of the second kind.  

The dimensionally 
regularized, thermal cosmological constant $\hat{\Lambda} (\beta)$
 manifestly satisfies the thermal duality relation $\beta 
\hat{\Lambda}(\beta) = \tilde{\beta} \hat{\Lambda}(\tilde{\beta})$ in 
full accordance with the thermal stability of modular invariance.  The 
thermal duality symmetry immediately yields the 
asymptotic formula as follows: $\hat{\Lambda} (\beta \sim 0) \sim 2\pi^2 
\alpha^\prime / \beta^2 \cdot \hat{\Lambda} (\beta^{-1} \rightarrow 0) = 
2\pi^2 \alpha^\prime / \beta^2 \cdot \Lambda$ for the heterotic 
thermal string theory, where $\Lambda$ 
literally reads the $D = 10$ zero-temperature, one-loop cosmological 
constant which is in turn guaranteed to vanish automatically as an 
inevitable consequence of the Jacobi identity $\theta_2^4 - \theta_3^4 + 
\theta_4^4 = 0$ for the theta functions. 
The present observation is 
paraphrased $\grave{a}\: la$ Osorio as follows:  The 
thermal duality symmetry is inherent to the fact that the total number 
of degrees of freedom vanishes at extremely high temperature $\beta 
\sim 0$ in the sense of the modular invariant counting.  Accordingly, 
it seems possible to claim $\grave{a}\: la$ Atick and Witten that 
the heterotic thermal string will be 
asymptotically described at high temperature by underlying 
topological theory. The present view may deserve more than passing 
consideration in an attempt 
to substantialize the geometrical ideas purely topological in 
character.  

Let us describe the singularity 
structure of the dimensionally 
regularized, dual symmetric thermal amplitude 
$\hat{\Lambda}(\beta)$.  The position of the singularity $\beta_{|p|, 
|q|}$ is determined by solving $\beta/\tilde{\beta} \cdot p^2 + 
\tilde{\beta}/\beta \cdot q^2 - 6 = 0$ for every allowed $(p, q)$ in 
eq.~(8).  Thus we obtain a set of solutions 
as follows: $\beta_{1, 1} = \beta_H = (\sqrt{2} + 
1) \pi \sqrt{2\alpha^\prime}$ and $\tilde{\beta}_{1, 1} = \tilde{\beta}_H 
= 
(\sqrt{2} - 1)\pi \sqrt{2\alpha^\prime}\:$ which form the leading 
branch points of the square root type at $\beta_H$ and $\tilde{\beta}_H$, 
respectively.     
We are now in the position to touch upon the global 
phase structure of the heterotic thermal string ensemble.  
There 
will then exist three phases in the sense of the thermal duality 
symmetry as follows: (i) 
the $\beta$ channel canonical phase in the tachyon-free region $(2 +
\sqrt{2})\pi \sqrt
{\alpha^\prime} = \beta_H \leq \beta < \infty$, (ii) the dual 
$\tilde{\beta}$ channel canonical phase in the tachyon-free region 
$0 < \beta \leq 
\tilde{\beta}_H = (2 - \sqrt{2})\pi \sqrt{\alpha^\prime}$ and 
(iii) the self-dual microcanonical phase in the tachyonic region 
$\tilde{\beta}_H < 
\beta 
< \beta_H$.  
There 
will appear no effective splitting of the microcanonical domain because of 
the absence of the self-dual leading branch point at $\beta_0 = \tilde{\beta}_0 
= \pi \sqrt{2\alpha^\prime}$.  As a 
consequence,  it still remains to be clarified whether the so-called maximum 
temperature of 
the heterotic string excitation is asymptotically described as
 $\beta_0^{-1} = \tilde{\beta}_0^{-1}$ in proper respect to the self-duality of 
 the microcanonical 
 phase.  
  
We have succeeded in shedding some light upon physical aspects of the thermal 
duality symmetry in full harmony with the thermal stability 
of modular 
invariance through the infrared 
behaviour of the one-loop cosmological constant for the dimensionally 
regularized, $D = 10$ heterotic thermal string theory based upon the 
TFD algorithm.

\end{document}